\documentclass{article}
\usepackage{cite}
\usepackage{amsmath,amssymb,amsfonts}
\usepackage{algorithmic}
\usepackage{graphicx}
\usepackage{textcomp}
\usepackage{breqn}
\usepackage{booktabs}
\usepackage{caption}
\usepackage{subcaption}
\usepackage{amssymb}
\usepackage{pifont}
\newcommand{\cmark}{\ding{51}}
\newcommand{\xmark}{\ding{55}}
\usepackage{bm}
\usepackage{multirow}
\usepackage{siunitx}

\begin{document}
\title{3D Tomographic Pattern Synthesis for Enhancing the Quantification of COVID-19}

\author{Siqi Liu$^\dagger$, Bogdan Georgescu$^\dagger$, Zhoubing Xu$^\dagger$, Youngjin Yoo$^\dagger$, \\Guillaume Chabin, Shikha Chaganti, Sasa Grbic, Sebastian Piat, \\Brian Teixeira, Abishek Balachandran, Vishwanath RS, Thomas Re, \\Dorin Comaniciu
\thanks{
$^\dagger$ indicates equal contribution.\newline
Siqi Liu, Bogdan Georgescu, Zhoubing Xu, Youngjin Yoo, Shikha Chaganti, Sasa Grbic, Sebastian Piat, Brian Teixeira, Dr. Thomas Re, Dorin Comaniciu are with Siemens Healthineers, Princeton, NJ, USA. (email: siqi.liu@siemens-healthineers.com) \newline\newline
Guillaume Chabin is with Siemens Healthineers, Paris, France.\newline\newline
Dr. Abishek Balachandran, Dr. Vishwanath RS are with with Siemens Healthineers, Bangalore, India.\newline\newline
We gratefully acknowledge the contributions of Health Time, Hospital Foch, The Methodist Hospital, Northwell Health, Universitaetsspital Basel, University of British Columbia, and multiple other frontline hospitals for our collaboration and for providing the data used in this study.\newline\newline
The  authors thank  the  National  Cancer  Institute  for  access  to  NCI’s  data collected  by  the  National  Lung  Screening  Trial  (NLST).  The  statements contained herein are solely those of the authors and do not represent or imply concurrence or endorsement by NCI.
\newline\newline
The authors also thank the COPDGene for providing the data. The COPDGene study (NCT00608764) was funded by NHLBI U01 HL089897 and U01 HL089856 and also supported by the COPD Foundation through contributions made to an Industry Advisory Committee comprised of AstraZeneca, BoehringerIngelheim, GlaxoSmithKline, Novartis, and Sunovion.
}
}
\date{}
\maketitle

\begin{abstract}
The Coronavirus Disease (COVID-19) has affected 1.8 million people and resulted in more than 110,000 deaths as of April 12, 2020 \cite{jhudashboard}. 
Several studies have shown that tomographic patterns seen on chest Computed Tomography (CT), such as ground-glass opacities, consolidations, and crazy paving pattern, are correlated with the disease severity and progression \cite{bernheim2020chest,chung2020ct,zhao2020relation,kanne2020essentials,guan2020clinical}. CT imaging can thus emerge as an important modality for the management of COVID-19 patients. 
AI-based solutions can be used to support CT based quantitative reporting and make reading efficient and reproducible if quantitative biomarkers, such as the Percentage of Opacity (PO),
can be automatically computed. However,
COVID-19 has posed unique challenges to the development of AI, specifically concerning the availability of appropriate image data and annotations at scale. 
In this paper, we propose to use synthetic datasets to augment an existing COVID-19 database  to tackle these challenges.
We train a Generative Adversarial Network (GAN) to inpaint COVID-19 related tomographic patterns on chest CTs from patients without infectious diseases.
Additionally, we leverage location priors derived from manually labeled COVID-19 chest CTs  patients to generate appropriate abnormality distributions.
Synthetic data are used to improve both lung segmentation and segmentation of COVID-19 patterns by adding 20\% of synthetic data to the real COVID-19 training data. 
We collected 2143 chest CTs, containing 327 COVID-19 positive cases, acquired from 12 sites across 7 countries.
By testing on 100 COVID-19 positive and 100 control cases, we show that synthetic data can help improve both lung segmentation (+6.02\% lesion inclusion rate) and abnormality segmentation (+2.78\% dice coefficient), leading to an overall more accurate PO computation (+2.82\% Pearson coefficient).
\end{abstract}

\section{Introduction}
\label{sec:introduction}

Coronavirus Disease 2019 or COVID-19 is a rapidly growing pandemic. As of April 12, 2020, more than 1.8 million people have been affected by the condition worldwide, resulting in over 110,000 deaths \cite{jhudashboard}. COVID-19 typically presents with fever, cough and dyspnea, although a significant percentage of people might be asymptomatic\cite{mizumoto2020estimating}. Estimates show that approximately 20\% of the patients will develop severe symptoms due to pneumonia\cite{guan2020clinical}. While there is no treatment available at the moment, severe cases require hospitalization for management of respiratory symptoms and other serious consequences such as multiple organ failures. The pandemic has caused a severe burden on healthcare systems around the world, leading to a wide range of issues from testing availability to limited hospital and ICU beds\cite{ji2020potential,emanuel2020fair}. The disease is confirmed by an RT-PCR test, which has a sensitivity as low as 70\%\cite{fang2020sensitivity,ai2020correlation} and might cause significant delays due to the lack of real-time testing. Recently, the Fleischner Society released recommendations for chest imaging, including CT and X-ray, for COVID-19\cite{rubin2020role}. In an environment with constrained resources, such as New York City and parts of Italy, Spain, Iran, and China, it is recommended that chest imaging be used for rapid triage and prioritization of patients in emergency rooms and hospitals. Even in situations with no serious resource constraints, it is recommended that chest imaging is performed for patients suspected of COVID-19 with moderate to severe symptoms to establish a baseline pulmonary status, perform risk stratification, and verify false negatives.
Several studies have shown that features seen on lung CT, namely ground-glass opacities (GGO), crazy paving patterns and consolidations, are correlated with disease severity and progression in COVID-19\cite{bernheim2020chest,chung2020ct,zhao2020relation,kanne2020essentials,guan2020clinical}. 
For each of the tasks of triage, severity assessment and progression tracking, the extent of abnormality must be measured quantitatively. An automated tool that can compute the extent and severity of CT abnormalities is therefore very useful. In this paper, we present an AI system for detection and quantification of abnormalities associated with COVID-19. We also present the unique challenges that arise due to the nature of the COVID-19 emergency and solutions to address each of these challenges.

Development of an AI system in an emerging outbreak presents several unique problems, as compared to developing a system for well-known diseases:
\begin{itemize}
\item \textbf{Data.} There are well-established and curated databases available for several pulmonary diseases such as lung cancer (NLST, LUNA, TCIA) and emphysema (COPDgene). However, COVID-19 is a new disease with emerging imaging protocols. Collection and curation of chest imaging data for COVID-19, therefore, provides new challenges. 
\newline \textit{Technical Challenges:} AI systems need a large amount of data for training, validation and testing. The data needs to be diverse, collected from multiple countries and hospitals, including several imaging devices to guarantee the generalizability of system across diverse patient populations and imaging devices. 
\item \textbf{Clinical or Domain Knowledge.} The clinical indications for the role of imaging has been evolving over the past few months, in parallel to the development of our project. Likewise, the clinical characterization and benchmark quantification of the disease is constantly developing. There are a variety of the CT patterns in the long-tail distribution, variations in the patient population, variation in the scale, intensity, texture, and location of pulmonary abnormalities. 
\newline \textit{Technical Challenges:} Ambiguities are present for annotation of data with respect of boundaries, inclusion criteria and quality control. Existing AI-dependent pipelines such as landmark detection and organ segmentation can be broken due to new and severe abnormalities. 
\item \textbf{Time Sensitivity.} There is an urgent need for technical solutions for evaluation of COVID-19 due to the enormous burden on public health, social and economic structures. 
\newline \textit{Technical Challenges:} There is limited time for iterative development of AI models with limited resources.
\end{itemize}

To tackle the challenges presented above we propose to generate synthesized data to support the training and development of abnormality detection and segmentation methods relevant for COVID-19. Additionally, the synthesized data can be leveraged to improve the lung segmentation models without requiring additional image labelling. In this paper, we propose to inpaint the main patterns of COVID-19, i.e., GGO, consolidations and crazy paving on to CT images from control patients. The data synthesis is trained as a 3D GAN that maps a masked real chest CT image to a chest CT image inpainted with COVID-19 abnormalities: GGO, consolidations and crazy paving patterns. We use randomly deformed 3D meshes and spatial probability map, derived from patients diagnosed with COVID-19, to synthesize the abnormality masks. 
\par
The synthetic data is mixed with real data to improve the performance of both the lung segmentation and the COVID-19 abnormality detection and segmentation. We constructed a database of chest CT scans from 327 COVID-19 positive patients from 12 clinical sites across 7 countries. Also, we added chest CT scans from 471 patients diagnosed with pneumonia patterns and 510 control patients to the database. 
By  evaluating  on  a  benchmark  data set containing  100  COVID-19  positive  cases  and  100  control patients,  we  show  that  the  synthetic  data  can  be  helpful  to improve  the  lesion  inclusion  rate  of  the  lung  segmentation by  6.02\%  and  the  precision  of  automatically  computed  PO by 3.75\% for 2D lesion segmentation network and 2.82\% for the 3D lesion segmentation network. 

\section{Related Work} 
\subsubsection{AI-aided evaluation for COVID-19}
There is limited work done so far in development and deployment of automated systems for the evaluation of COVID-19. Shan et al showed that a deep learning-based system can be used to detect and measure the severity of COVID-19 CT abnormalities\cite{shan+2020lung}. In this work, they used a V-net with a bottle-neck structure for segmentation of regions affected by COVID-19. Li et al show that COVID-19 can be differentiated from other types of pneumonia and lung diseases with the help of a CNN-based classifier\cite{li2020artificial}. While several papers also show varying levels of success in classifying COVID-19 vs controls using deep learning methods \cite{wang2020deep,xu2020deep,gozes2020rapid}, the role of CT imaging in diagnosis and screening remains debatable since a percent of patients who test positive for COVID-19 do not show CT changes. Population studies involving broad testing revealed that out of all patients who tested positive, 46\% of the asymptomatic patients \cite{inui2020chest}, 15\% of the patients with mild symptoms and 5\% of the patients with severe symptoms show no changes on CT. However, the extent of CT abnormalities, when they are present, are associated with severity and progression \cite{bernheim2020chest,chung2020ct}. Therefore, in this paper we focus on the quantification of abnormalities associated with COVID-19.

\subsubsection{Image synthesis based data augmentation in medical image analysis}

There has been an increasing interest in synthesizing objects in medical images to augment existing training set for better diversity due to advances in advances of generative deep learning models in recent years.
Many recent studies proposed to use generative networks for either transferring the image modalities \cite{nie2017medical,chartsias2017adversarial,wang2018unsupervised,yang2018unpaired} or inject synthesize objects \cite{yang2019,liu2018decompose,jin2018ct,xu2019tunable,xu2019correlation,gao2019augmenting,han2019synthesizing,wang2019wgan,liu_adversarial,shin2018medical,frid2018gan} to improve the performance of diverse medical imaging related AI applications.
It was observed in \cite{liu2018decompose} that only adding the hard synthetic cases into the training set could improve the supervised model performance with a large margin since the majority of the synthetic samples would add little values since they can be successfully recognized by a network that is trained on a large-scale dataset. Thus, authors of \cite{liu2018decompose} use both the discriminator error and the classification error to select only the hard synthetic cases to be added to the augmented dataset.
In \cite{liu_adversarial}, authors proposed to synthesize hard samples by drawing latent code from a fully differentiable synthesizer using projected gradient descent. In this work, we use a similar hard-case sampling approach as \cite{liu2018decompose} since the locations of the synthetic tomographic patterns are not yet fully differentiable.

\section{Methods}

\begin{figure}
    \includegraphics[width=0.9\linewidth]{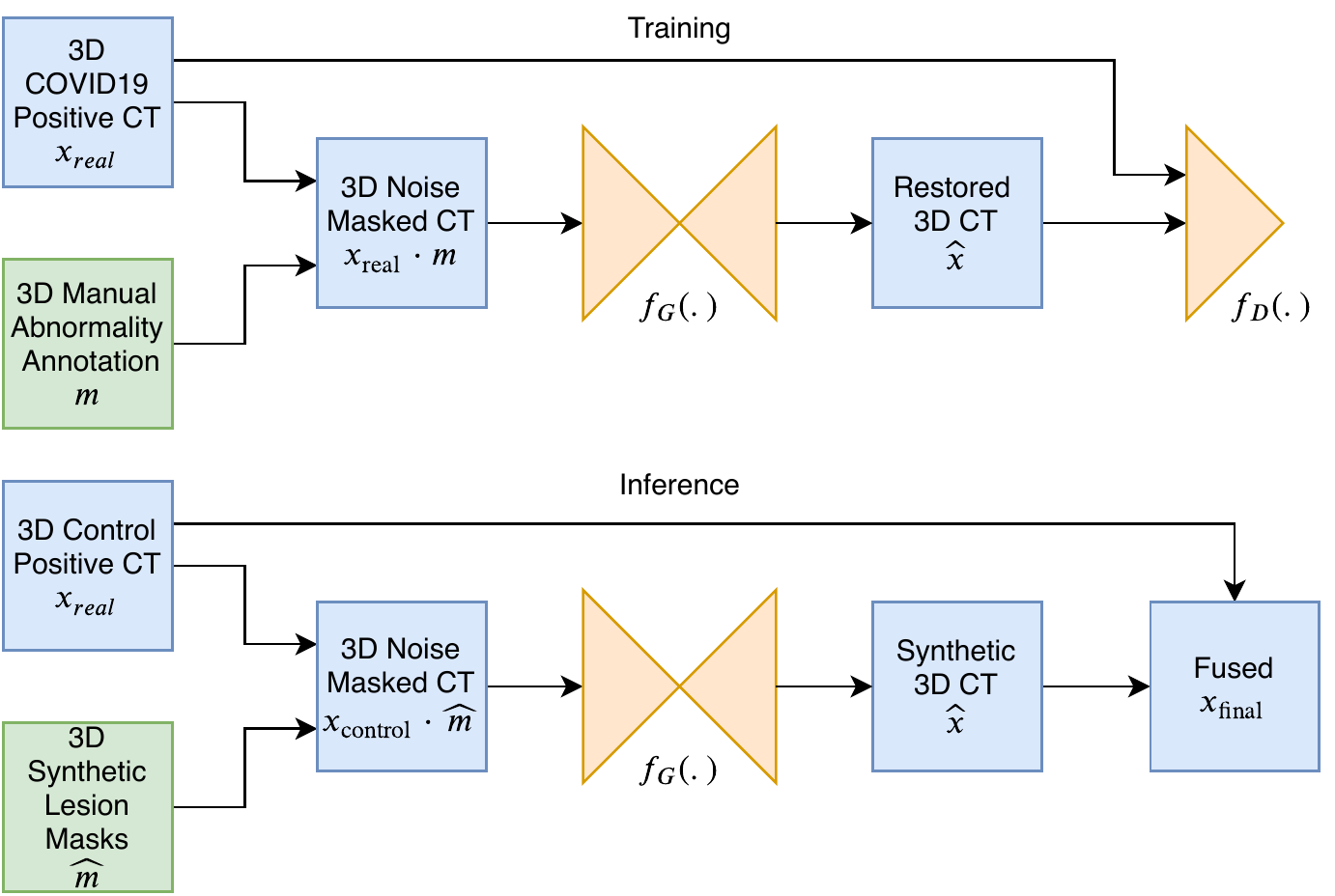}
    \label{fig:syn-overall}
    \caption{The data-flow of the training and the inference of the proposed tomographic pattern synthesis framework.}
\end{figure}

\subsection{Synthesis of Tomographic COVID-19 Patterns}
Given 3D CT images $I$ and the annotated segmentation masks of COVID-19 opacities $M$,
we train a GAN generator to obtain the mapping $I = f(I \cdot M)$ to synthesize COVID-19 opacities on arbitrary CT images where $\cdot$ denotes the filling uniform noise in the masked region. We use a heuristic approach to synthesize random 3D masks during inference. To avoid intensity bias of the control lung tissues, the output synthetic images are fused with the original images using the synthetic masks.

\subsubsection{CT Image Synthesis} %

The 3D chest CT images are re-sampled to the resolution $0.75\times0.75\times1mm$. The image intensities are normalized to $[-1, 1]$ using the standard lung window with level -600 and window width 1500.
We fill uniform noise with values between $[-1, 1]$ into the image regions annotated with COVID-19 abnormality patterns thus the patterns are hidden from the generator network.
The training 3D patches are cropped from the normalized images with size $384\times384\times18$ each and are ensured to be centered regarding both lungs. To train the synthesizer we only keep the positive training patches with annotated COVID-19 patterns.

The generator $f_G$ is built with a 3D UNet \cite{ronneberger2015u}. For each building block of the generator, we use the instance normalization \cite{instancenorm} followed by a $3\times3\times3$ convolution layer and LeaklyReLU. The final generator output has the same size as the input and is activated with the Tanh function.
The discriminator network $f_D$ is built with a simple multi-layer CNN. We use the Spectral Normalization \cite{sn} in the discriminator to balance the learning speed of both networks.
For both the real and the fake input to the discriminator, we add a 3D tensor $n \sim \mathcal{N}(0, 0.2)$ drawn from the Gaussian noise to avoid the discriminator from pre-maturing during the early iterations. The noise biased inputs are clipped back to $[-1, 1]$ before being fed to the network.

The objectives for training the synthesizer can be summarized as following
\begin{equation}
    \hat{x}_{fake} = f_{G} (x_{real} \cdot m)
\end{equation}
\begin{equation}
\begin{split}
     L_{D} = & \|f_D(n + x_{real}) - t(1))\|^2_2 + \\ 
     & \|f_D(n + \hat{x}_{fake}) - t(0))\|^2_2
\end{split}
\end{equation}
\begin{equation}
\begin{split}
    L_{G} = & \lambda_1 |x_{fake} \circ \neg m - \hat{x}_{fake} \circ \neg m| + \\
    & \lambda_2 |x_{fake} \circ m - \hat{x}_{fake} \circ m | - \\
    & L_D
\end{split}
\end{equation}
where $x_{real}$ is the real 3D CT training patch; m is the annotated mask with COVID-19 patterns; $\cdot$ denotes of operation of filling the uniform noise into the mask regions. 
$t(.)$ is a target tensor filled with a constant value (0 or 1) with the same size as the discriminator output. We use the LSGAN objective which measures the L2 errors between the discriminator output and the target. $\circ$ denotes tensor element-wise multiplication. $\neg m$ is the reversed mask that covers the non-impacted areas. $\lambda_1$ and $\lambda_2$ are two hyper-parameters to balance the L1 losses in the COVID-19 affected areas as well as the weight of the discriminator loss. In all of our experiments, we fixed $\lambda_1 = \lambda_2 = 10$. We use ADAM to optimize both networks with the learning rate of $0.001$ for generator and $0.004$ for the discriminator.
In our experiments, we trained a general COVID-19 generator and a consolidation biased generator.
The former is trained with all the annotated COVID19 patterns.
The latter is finetuned based on the former with all the training patches having above $-200$ mean intensity in the annotated regions.

\begin{figure}[!htb]
    \centering
    \includegraphics[width=1.0\linewidth]{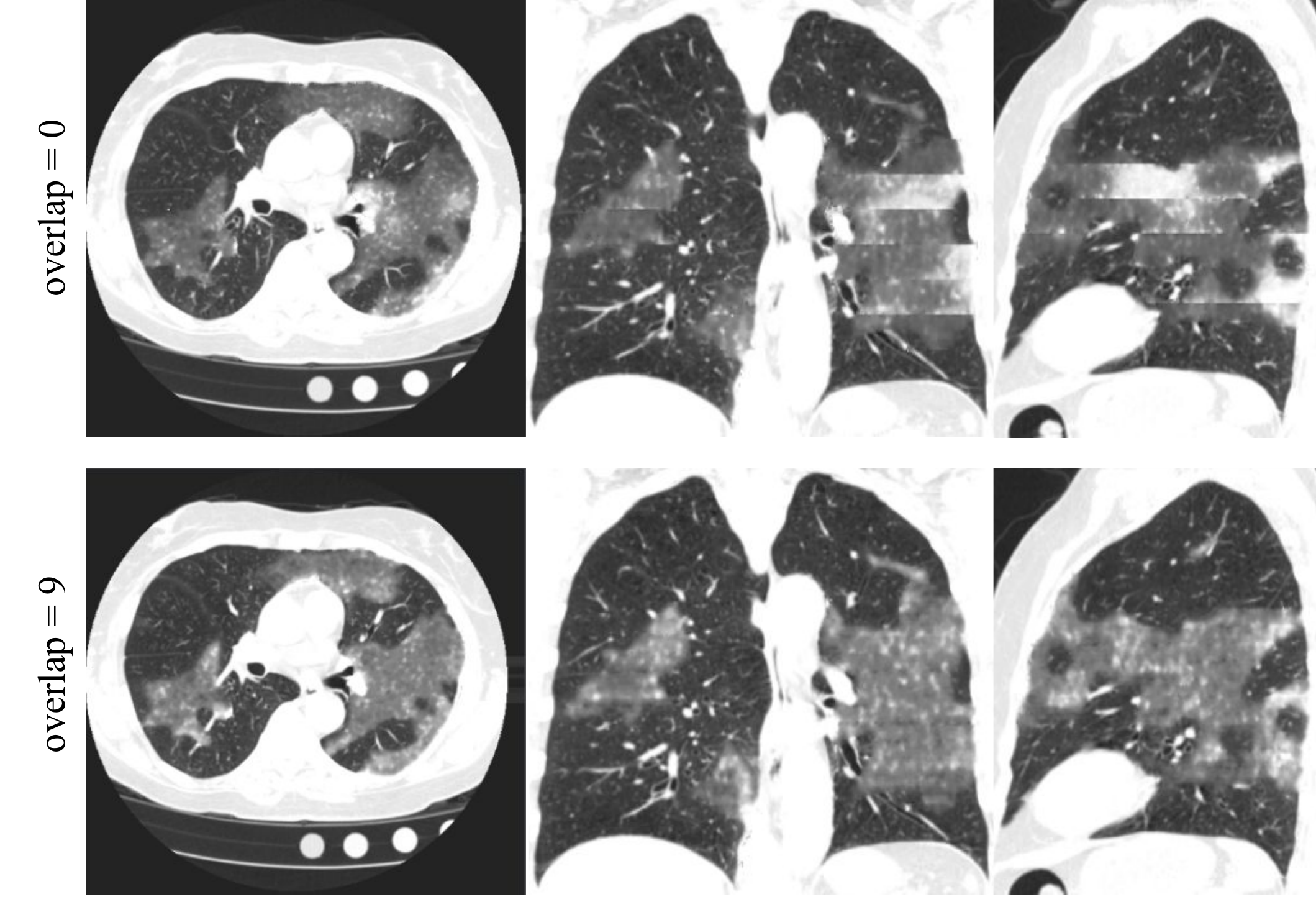}
    \caption{The example slices of the model output with input overlap = 0 and overlap = 9. From left to right are the example slices of axial, coronal and sagittal views. The 3D volume predicted with overlap = 9 shows better consistency along the z dimension.}
    \label{fig:overlap}
\end{figure}

To generate synthetic COVID-19 patterns on arbitrary CT images without manual segmentation masks are available, we use the synthetic COVID-19 segmentation masks $\hat{m}$ as described in Sec.~\ref{sec:masksyn} and mask the image with the uniform noise injected.
The same scaling, cropping, and intensity normalization is performed as in training.
We predict the output volume in a sliding window approach by only moving along the z dimension. The window size is fixed as the same as the training blocks $384\times384\times18$.
There is an overlap of 9 between every two steps. The overlapped region of the generator input is filled with the output from the previous step to avoid the discontinuity artefacts as shown in Fig.~\ref{fig:overlap}. Thus except the first step, generator predictions are conditioned on the previous generator outputs.

\begin{figure*}
    \centering
    \includegraphics[width=1.0\linewidth]{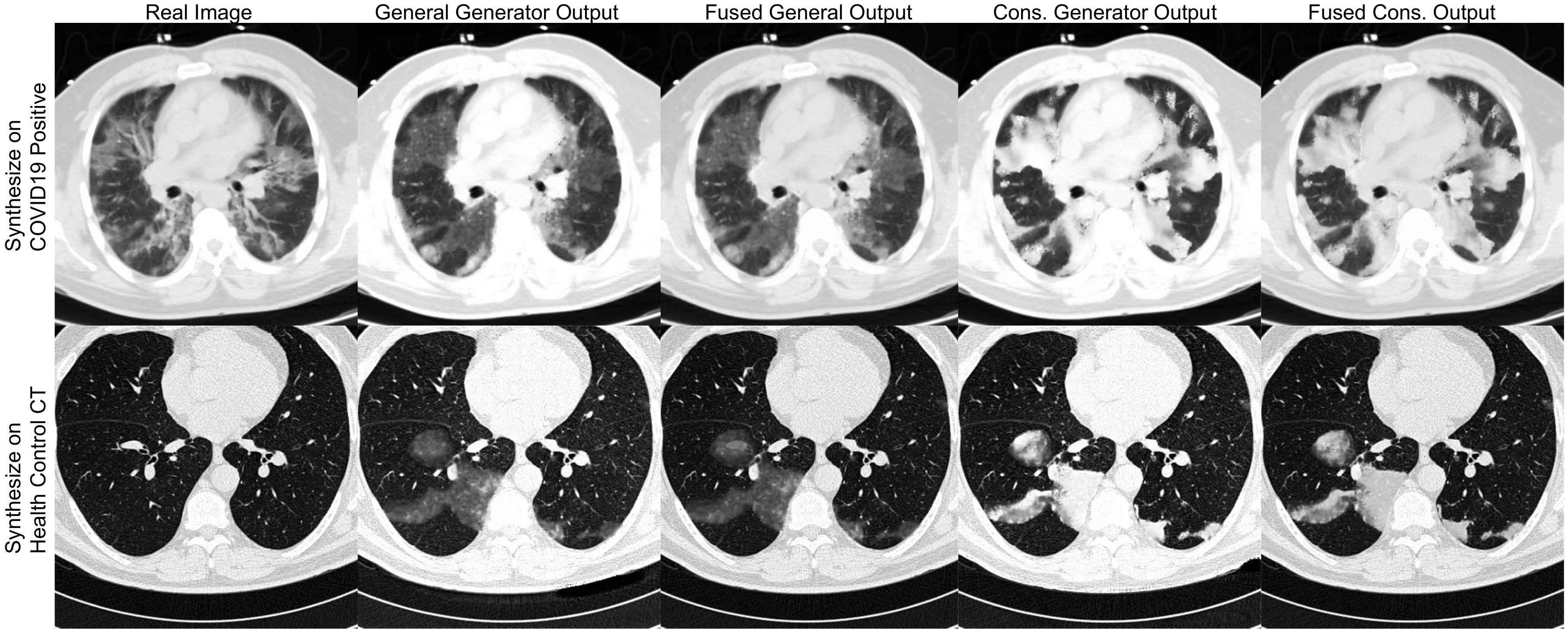}
    \caption{Example axial slices to illustrate the visual difference in different synthesis settings. From left to right: the original image; the model output from the generator trained with all the COVID-19 positive images; The output of fusing the generator output in the second column with the original image; The output of the generator trained with only high-intensity abnormalities; The output of fusing the model in the fourth column and the original image. The first row is obtained by inpainting the patterns on-to a COVID-19 patient using the same mask manually annotated. The second row is obtained by generating the COVID-19 patterns on-to a control CT using a synthetic mask.}
    \label{fig:model_output}
\end{figure*}

We only extract the synthetic abnormality regions from the generator output and fuse them to the original images since (1) there could be an intensity bias as shown in Fig. \ref{fig:model_output} and (2) we cannot guarantee that no other abnormality patterns hallucinated by the generator as observed in \cite{cohen2018distribution} (3) ground glass opacities are semi transparent in nature and hence the underlying vessels and bronchi are visible through these opacities.
We first blend the output with the original image with a weighted sum as
\begin{equation}
    x_{blend} = \beta \alpha \hat{x} + (1 - \beta) x_{control}
\end{equation}
where $\beta$ is the constant weight for the synthetic image. We use $\alpha$ to adjust the intensity of the generated abnormality for areas above -200 HU. The blended image $x_{blend}$ is then fused to the original image by cropping using the abnormality mask $\hat{m}_{smooth}$. The mask boundary is smoothed using a linear distance transform. The output of both models are shown in Fig.~\ref{fig:model_output}.
\begin{equation}
    x_{final} = x_{control} \circ \neg \hat{m}_{smooth} + x_{blend} \circ \hat{m}_{smooth}
\end{equation}

\subsubsection{Mask Synthesis} 
\label{sec:masksyn}

Our mask synthesis algorithm operates on 3D meshes, giving full
control of the geometry of the synthetic abnormality and ensuring closed shape. Starting from an initial template closed sphere, we randomly
select $N$ points on its surface. Then, for each point, we apply an
affine transformation function of a random amplitude factor $\lambda$. Such transformation is propagated to neighbouring vertices defined by a
distance threshold of $\delta$. Thus, for each sampled vertex $v_i$ and each neighbor vertex $n_j$, the affine factor $\alpha_j$ is defined as:

$$ \alpha_j = 1 + ((\delta - |v_i - n_j|) * \lambda_i) $$

Additionally, we apply a Laplacian smoothing followed by Humphrey filtering, as proposed in \cite{improved_laplacian}. Finally, we rasterize the resulting mesh to generate a 3D mask using recursive subdivision, as proposed in \cite{voxelize}.

\begin{figure}
    \centering
    \includegraphics[width=\linewidth]{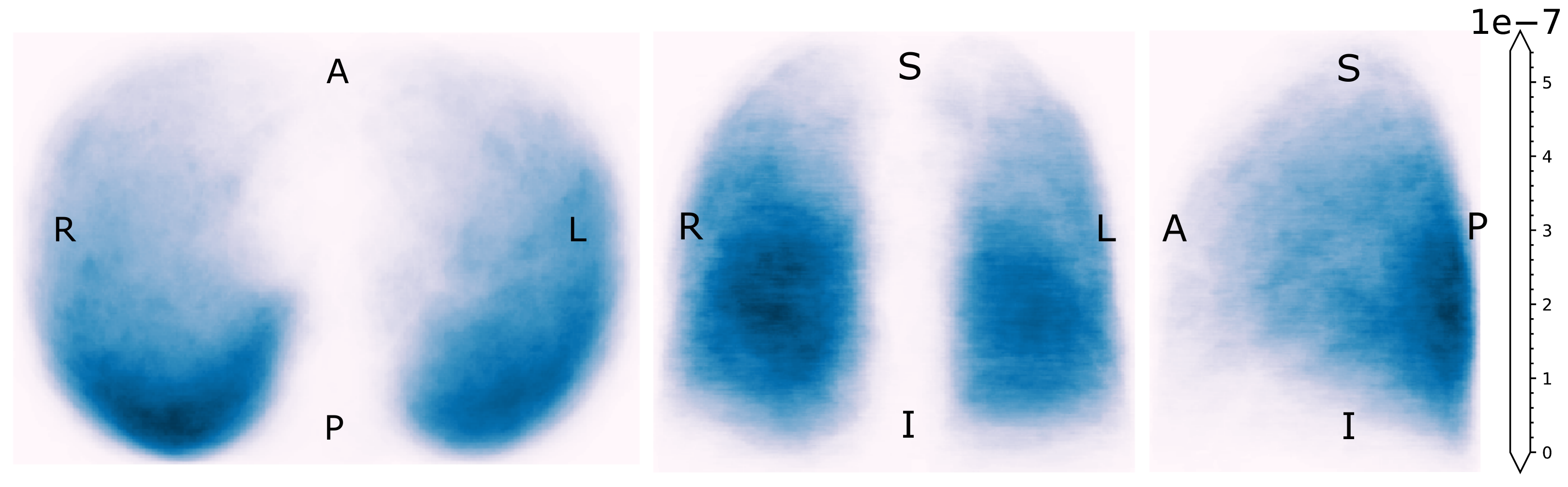}
    \caption{The spatial distribution probability map generated using the COVID-19 tomographic pattern annotations of the training cases. It is used for sampling the locations of the synthetic patterns.}
    \label{fig:spatial_prob}
\end{figure}

\begin{figure}[htb!]
     \centering
     \begin{subfigure}[b]{0.45\linewidth}
         \centering
         \includegraphics[width=\linewidth]{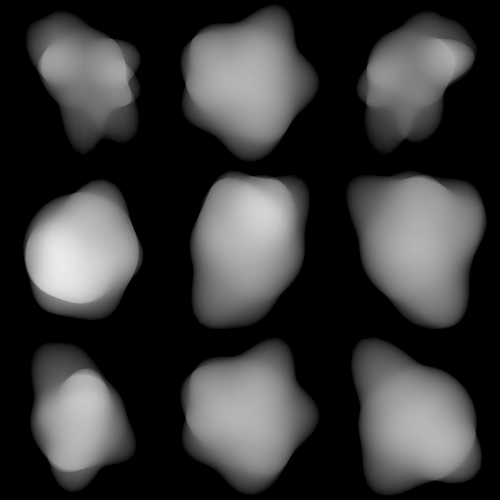}
         \caption{$N=10, \lambda=1.5$}
         \label{fig:mesh1}
     \end{subfigure}
     \hfill
     \begin{subfigure}[b]{0.45\linewidth}
         \centering
         \includegraphics[width=\linewidth]{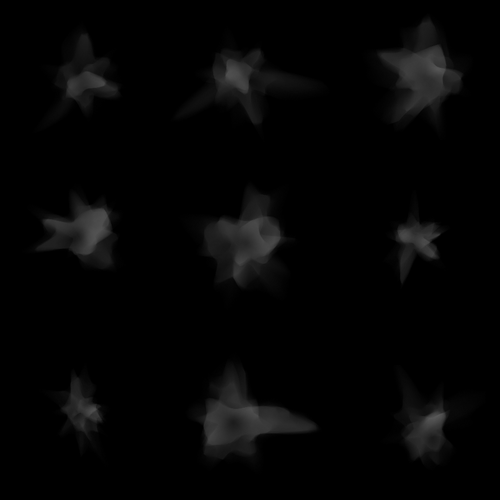}
         \caption{$N=200, \lambda=2.5$}
         \label{fig:mesh2}
     \end{subfigure}
\caption{Example 3D synthetic lesion masks randomly generated with different parameters.}
\label{fig:meshes}
\end{figure}

\begin{figure*}
    \centering
    \includegraphics[width=1.0\linewidth]{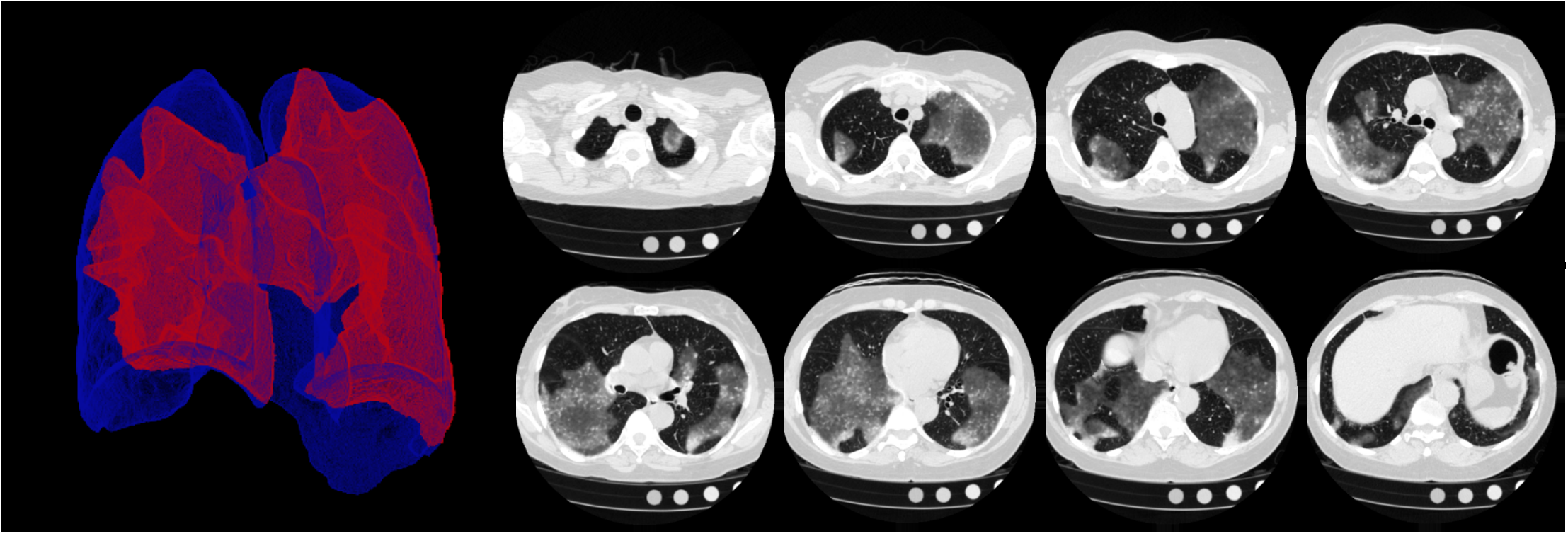}
    \caption{Left: The combined 3D synthetic mask showed with the automatically computed lung segmentation. Right: The example axial slices drawn from the synthetic 3D volume generated using the synthetic mask.}
    \label{fig:syn_mask}
\end{figure*}

The mask of each abnormality connected component is first synthesized independently as shown in Fig. \ref{fig:meshes}. We then combine them together by taking the union of all the masks. 
It is known that COVID-19 usually presents with opacities in subpleural, peripheral, bilateral and multilobar locations.
To simulate the spatial distribution of the COVID-19 tomographic patterns, we compute a spatial probability map using the aligned manual annotations of the COVID-19 positive cases as Fig.~\ref{fig:spatial_prob}. We sample the lesion center locations from the probability map and then map the sampled locations to the corresponding image space of each image. The combined mask is cropped using the automatically computed lung mask as shown in Fig.~\ref{fig:syn_mask}.

\subsection{Sythesis augmented lung segmentation} 
\label{sec:lung_seg_methods}

During an initial evaluation, a pre-trained DI2IN \cite{yang2017automatic} on 8006 chest CT data sets (no severe pneumonia patterns) had difficulty to capture the affected areas on COVID-19 cases, especially heavy consolidation at the posterior bottom and periphery of the lungs. This directed our investigation to focus on the affected areas. To start with, 675 and 60 images with moderate pneumonia patterns are included and annotated for training and validation respectively. The inclusion of pneumonia cases is helpful, but not sufficient due to the lack of heavy consolidation that is typical in COVID-19 cases. Therefore, 1530 images with synthetic consolidation patterns are further included for augmentation. Meanwhile, a few adjustments during training are applied to adjust for the inclusion of synthetic data. First, a weighted cross entropy is used to focus on high intensity areas. Consider $x$ the network input normalized from a CT image by the center of -624 Hounsfield Unit (HU) and the width of 1500 HU and clipped to the range of $[-1, 1]$, and $p$ the prediction out of the segmentation network $f_S(x)$, i.e., $ p = f_S(x) $., a voxel-wise weighted binary cross entropy is used that assigns additional attention on high intensity areas inside the lung. 
\begin{equation}
    L_S = - w[ylog(p) + (1 - y)log(1-p)]
\end{equation}
\begin{equation}
    w = 1 + \gamma_1 \frac{y}{1 + exp(-\gamma_2 x)}
\end{equation}
where $\gamma_1$ and $\gamma_2$ represents the magnitude and steepness of high intensity adjustment. Second, we remove the last skip connection (at the input size level) from DI2IN to constrain the lung shape despite the presence of the severe consolidation. 

The training process takes $128\times128\times128$ patches randomly sampled from $2\times2\times2 mm^3$ resampled volumes, and is driven by a learning rate of 0.001 using the ADAM optimization. The model for the epoch with the best performance on the validation set is selected.

\subsection{Synthesis augmented lesion segmentation}

To perform COVID-19 related abnormality segmentation, we use encoder-decoder based CNN architectures. To learn the relevant COVID-19 patterns, we train them on a training dataset from COVID-19, viral pneumonia and other interstitial lung diseases. To analyze the impact of synthetic COVID-19 data for training, we add them to the training dataset. We utilize both a 2D CNN approach and a 3D CNN approach. The 2D CNN approach aims to learn high-resolution in-plane image features by taking three axial slices as input to the network. We use the 3D CNN approach to efficiently model 3D context with anisotropic image resolution.  
\subsubsection{2D} %
The 2D CNN approach follows the U-Net architecture~\cite{ronneberger2015u} with an encoder to model COVID-19 relevant image features and a decoder to generate the segmentation mask. We employ the ResNet-32 architecture~\cite{he2016deep}, in which the feature encoder uses 5 ResNet blocks consisting of two $3\times3$ convolutions with batch normalization~\cite{ioffe2015batch} and ReLU~\cite{nair2010rectified}, followed by additive identity skip connection. The decoder has the same number of convolution blocks as in the encoder. The input to each decoding block is concatenated with the encoding features with the same resolution. The training images are resampled to have the in-plane resolution $0.6\times0.6$ mm. Then we compute the geometric center and crop the images with a fixed bounding box of size $512 \times 512$. We keep the original out-plane resolution and dimension. The images are clipped by the lung window with the width 1174 HU and level -150 HU, and then normalized to [-1,1]. The network is trained with Adam with decoupled weight decay regularization~\cite{loshchilov2017decoupled}. A soft dice loss~\cite{milletari2016v} is applied to the decoder output prediction to penalize the difference from ground-truth COVID-19 annotation during training. For data augmentation, we apply a random mirror flip for in-plane orientations with a probability of 0.5 and in-plane random translations that are limited to 10 voxels in each dimension. We perturb the image intensity within a random interval $[-10,10]$ HU.

\subsubsection{3D} %
Similar to the network architecture used in \cite{our_last_covid_paper}, we use a variant of the 3D U-Net network \cite{ronneberger2015u} with dense-convolutional blocks \cite{densenet} and anisotropic feature computation for higher resolution features and isotropic for lower resolution. Input CT volumes are pre-processed by resampling them to 1x1x3mm resolution and cropped based on the lung segmentation to a fixed $384\times384\times128$ box. Input data is masked by the lung segmentation and normalized using a standard lung window with width 1500HU and level $-600$ HU and clipped to $[0, 1]$. During training, additional data augmentation is performed by random intensity perturbation within a $[-20, 20]$ HU interval and random flipping along $x$ or $y$ directions. The 3D neural network uses convolutional blocks containing either $1\times3\times3$ or $3\times3\times3$ CNN kernels in dense blocks of convolution-BatchNormalization-LeakyReLU layers. For downsampling the encoder features are computed using a $1\times2\times2$ or $2\times2\times2$ convolution layers with a $1\times2\times2$ or $2\times2\times2$ stride and for upsampling transpose-convolution layers are used with same kernel sizes. The top two decoder-encoder network levels are using anisotropic features followed by three isotropic levels. The input to each decoder block is obtained by concatenating the corresponding encoder output features with the same resolution with the output of the previous upsampling block. The final output is using a softmax activation layer.
The 3D network is trained using the AdaBound optimizer \cite{adabound} which adaptively combines the Adam optimizer with SGD for faster convergence. We use the Jaccard index as the training loss function which we found that has stable behavior for imbalanced labels.

Among the advantages of using a 3D architecture is the ability to use 3D context to deal with in-plane partial volume effects as well as global lung context. Disadvantages include higher computational complexity and potentially higher complexity and overfitting in training due to a lower number of total samples. The choice of using anisotropic features is made as a compromise between computation complexity and having reasonable high-resolution features computed in the axial acquisition planes. 

\section{Data}

The lung segmentation algorithm and the lesion segmentation algorithm were developed and evaluated with the dataset summarized in Table.~\ref{tbl:data}. Please note that not all of the statistics are available due to the various levels of anonymization. The lung segmentation and the lesions segmentation were trained on dedicated datasets to address the challenges specific to each training task. 
The performance of the system was evaluated using the same testing set.

The testing set constitutes 100 control images and 100 COVID-19 positive images. The control group was randomly sampled from the non-pathological images in NLST \cite{nlst}. Candidates were identified from the clinical reports and visually confirmed by a trained user after selection. The 100 COVID-19 positive patients were sampled from data sources with a clinical confirmation. 110 candidates scans were randomly selected from 2 European and 2 American institutions. Ten datasets with the lowest Percentage of Opacity (PO) measured using the ground truth annotations were excluded. 
All volumes referenced to the patients selected in the testing set were excluded from any training sets.
The lesion segmentation training set constitutes the remaining 227 COVID positives cases collected from 10 clinical collaborators, augmented with 174 3D Chest CTs with pneumonia patterns, 297 cases with interstitial lung diseases.
The lung segmentation training set is made of 735 CT scans with both pathological (including pneumonia, interstitial lung disease) and control volumes. Please note that 187 datasets were common to the lesion segmentation training set and the lung segmentation training. 
The synthetic images used in this study are generated based on 510 control images acquired from the COPDGene cohort \cite{copdgene}. We synthesized 3 images based on each real control image resulting in 1530 synthetic images in total.

\begin{table}[ht]
\footnotesize
\caption{Properties of training and testing data used for lung segmentation and COVID-19 lesion segmentation. IQR: Interquartile rage.}
\centering
\resizebox{1\linewidth}{!}{
\begin{tabular}{p{0.09\textwidth}  || p{0.19\textwidth} || p{0.16\textwidth} || p{0.24\textwidth} || p{0.18\textwidth} }
\toprule
 & Lung Training & Synthetic Data Input  & Lesion Training & Testing  \\ \midrule
Datasets & Total: 735, Pneumonia: 30, ILD: 705 & Control: 510 & Total: 698, COVID: 227, Pneumonia: 174, ILD: 297 &  Total: 200, COVID-19: 100, Control: 100  \\ \hline
Data Origin &  Multiple sites including sites in USA, and Belarus & Multiple sites including sites in USA & Multiple sites including sites in USA, Canada, Spain, Switzerland, France, Czech Republic, and Germany & Multiple sites including sites in USA, Spain and Czech Republic \\ \hline
Sex &  Female: 136, Male: 208, Unknown: 391 & Female: 84, Male: 136, Unknown: 290 & Female: 211 , Male: 232, Unknown: 257 & Female: 71, Male: 112, Unknown: 17 \\
\hline
Age (years) &  Median: 64, IQR: 55.5-70 , Unknown: 231 & Median: 60, IQR: 53.5-69, Unknown: 367 & Median: 60, IQR: 54-67, Unknown: 425  & Median: 61, IQR: 56- 62.25, Unknown: 64 \\
\hline
Scanner Manufacturer &  GE: 257, Siemens: 94, Philips: 67, Toshiba: 1, Other/Unknown: 316 & Siemens: 268, GE: 167, Philips: 14, Unknown: 61 & Siemens: 349, GE: 203, Philips: 28, Toshiba: 10, Other/Unknown: 110 & Siemens: 62, GE: 68, Philips: 24, Toshiba: 28, Other/Unknown: 18\\
\hline
Slice Thickness [mm] &  $\leq 1.5$: 384 ; $(1.5, 3.0]$: 230 ; $> 3.0$: 121  &  $\leq 1.5$: 510, $(1.5, 3.0]$: 0, $> 3.0$: 0  & $\leq 1.5$: 381, $(1.5, 3.0]$: 239 , $> 3.0$: 78 & $\leq 1.5$: 56, $(1.5, 3.0]$: 124, $> 3.0$: 20 \\
\hline
Recon. Kernel& Soft: 136, Hard: 321, Unknown: 278 & Soft: 510, Hard: 0, Unknown: 0 &  Soft: 94, Hard: 346, Unknown: 258 & Soft: 119, Hard: 81, Unknown: 0\\
\bottomrule
\end{tabular}
}
\label{tbl:data}
\end{table}

The original data formats were either DICOM images or 3D Meta-Images. The 3D CT series were reconstructed from DICOM images by keeping the original resolution and reorienting the volume axially. 
The annotation of the data has been formalized as two independent tasks: the annotation of lungs and the annotation of lesions (COVID-19 opacities, pneumonia and interstitial lung disease).

\subsection{Lung Segmentation Training}

The ground-truth for each training data set was generated by expert users with a custom annotation tool. The user could load anonymized 3D CT series (volume), interact with the image (including 3 multi-planar reformatted images), draw and edit contours and mark regions with a pre-specified label for the lungs. The final mask was saved as a file together with the reference to the original anonymized CT series. The annotations were reviewed according to internal quality guidelines. Each annotation was reviewed by a second, more experienced user.

\subsection{Synthesis Augmented Lesion Segmentation Training}

The ground-truth for each training data set was generated by expert users with a 3D editing tool. The user could load anonymized 3D CT series (volume) and if provided, a pre-computed mask to initialize the annotation. The annotator would then edit the mask and mark abnormalities such as GGO, consolidation and crazy paving with a pre-specified label. The final mask was saved as a file together with the reference to the original anonymized CT series. The annotations were reviewed according to internal quality guidelines. Each annotation was reviewed by a board certified radiologist. The pre-computed masks were produced by previously trained networks. 
Only cases a priori identified as lesion-positive were sent for annotation.

\subsection{Synthesis Augmented Lesion Segmentation Testing}
The ground-truth was generated using the same approach as for the training data. 
In order to perform an inter-rater variability study, 13 random chest CT data sets from patients diagnosed with COVID-19 were given to two clinical experts for manual annotations. These 13 cases were randomly selected from our testing data set of COVID-19 positive patients.

\section{Results}

\subsection{Lung Segmentation}

\begin{figure}[!htb]
    \centering
    \includegraphics[width=\linewidth]{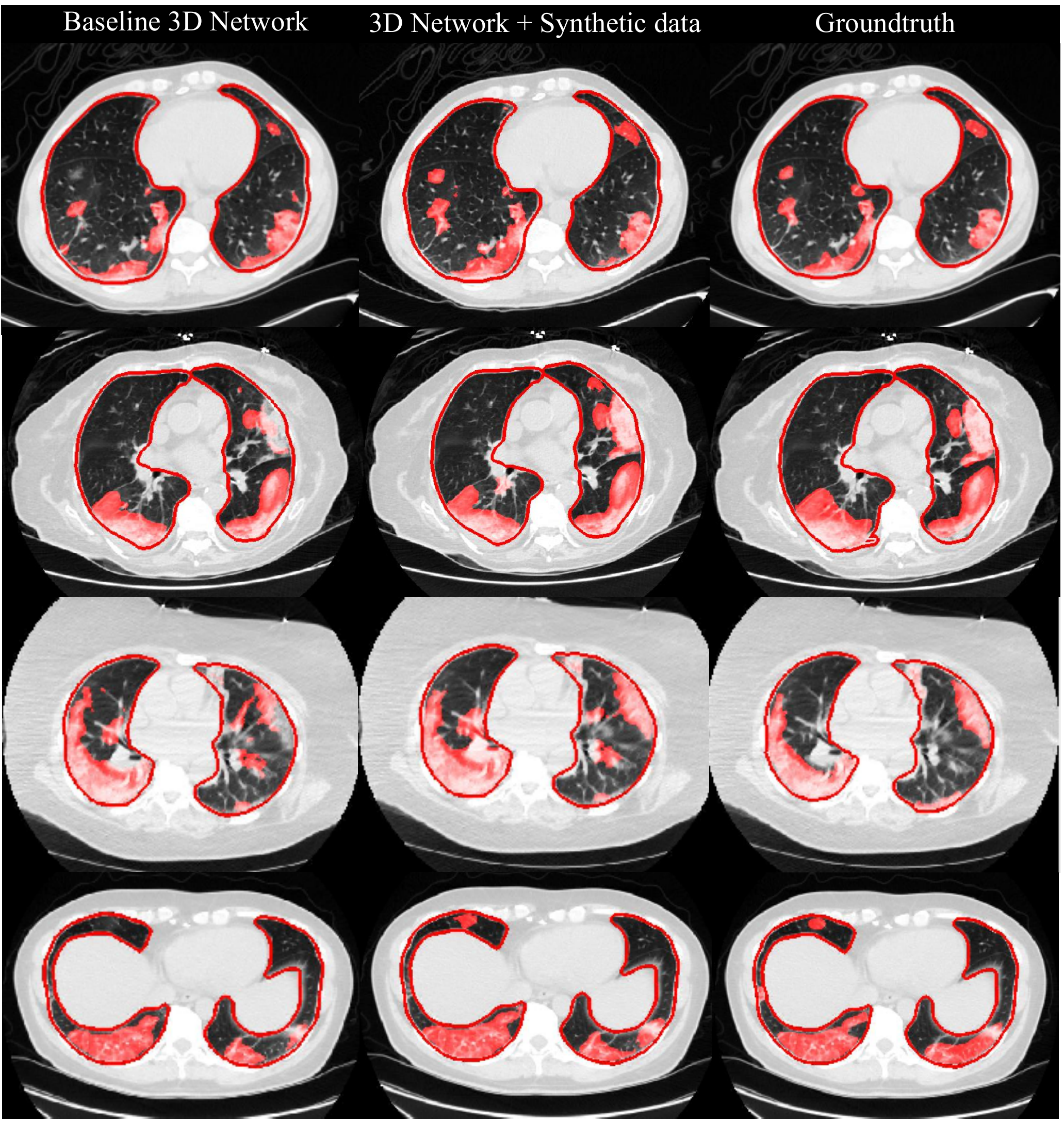}
    \caption{Example slices of the lesion segmentation network output overlaid on the CT axial slices.}
    \label{fig:covid-lesion}
\end{figure}

The lung segmentation is a prerequisite for lesion segmentation. Given this, it is crucial to have the abnormality region fully covered by the lung segmentation. The performance of different segmentation methods for the inclusion of abnormalities in the lung mask was not captured by the  traditional metrics like Dice similarity coefficient and average surface distance 
Therefore, we introduce a new metric called the lesion inclusion rate, i.e., $ LIR = |S_{lesion} \cap S_{lung}| / |S_{lesion}| $. The $LIR$ is computed for three lung segmentation methods, (a) one only trained with non-pneumonia data, (b) one fine-tuned with pneumonia data, and (c) one trained with both pneumonia data and COVID-like synthetic data along with some tailored adjustments described in section II-B. 

\begin{figure*}[!htb]
    \centering
    \includegraphics[width=1.0\linewidth]{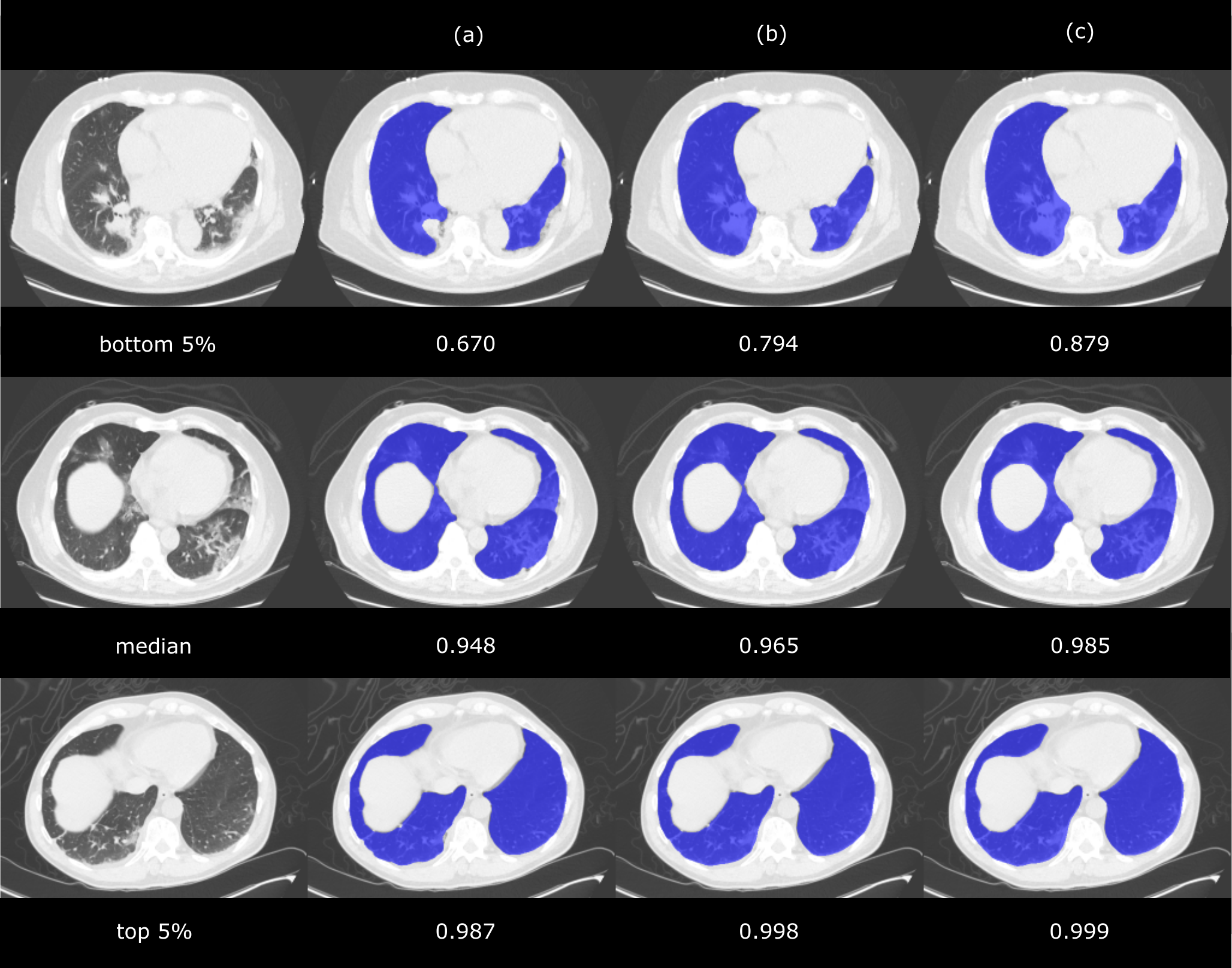}
    \caption{Qualitative results of lung segmentation on example cases with bottom 5\%, median, and top 5\% lesion inclusion rate among three methods (a) trained with non-pneumonia images, (b) finetuned on pneumonia images, and (c) proposed method trained on pneumonia images and synthetic images.}
    \label{fig:lung_qualitaive}
\end{figure*}

\begin{figure}
    \centering
    \includegraphics[width=0.5\linewidth]{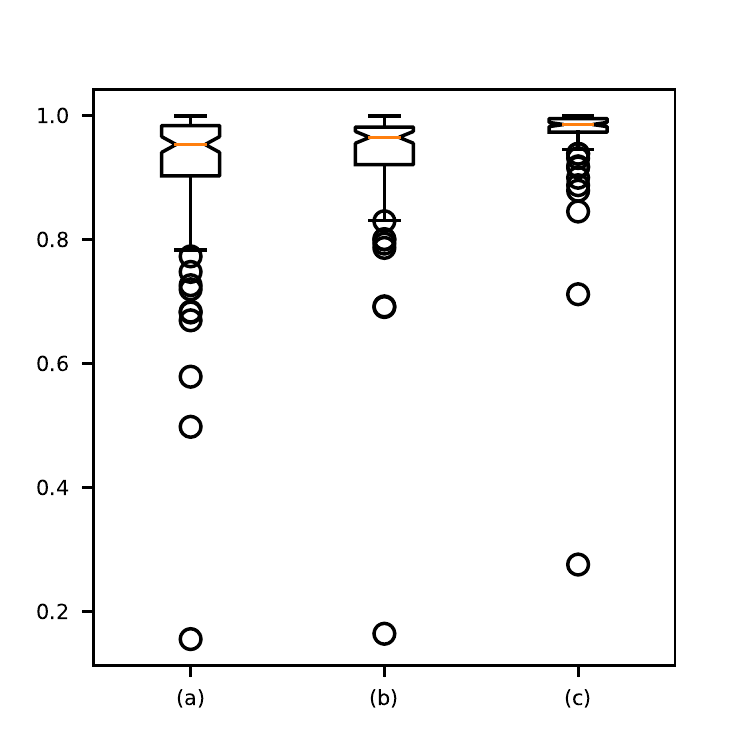}
    \caption{The lesion inclusion rate of the lung segmentation across three methods (a) trained with non-pneumonia images, (b) fine-tuned on pneumonia images, and (c) proposed method trained on pneumonia images and synthetic images}
    \label{fig:lung_ab_inclusion}
\end{figure}

From both qualitative (Fig.~\ref{fig:lung_qualitaive}) and quantitative (Fig.~\ref{fig:lung_ab_inclusion}) results, it can be seen that the methods trained with high abnormality data demonstrate better robustness in covering the lung regions with COVID-19 patterns. Without explicitly training on COVID-19 cases, an average LIR of 0.968 is achieved across 100 COVID positive cases through COVID pattern synthesis and associated model training adjustments, compared to an average LIR of 0.913 from a baseline method previously trained over 8000 images.

\subsection{COVID19 Tomographic Pattern Quantification}
We show example slices of the lesion segmentation network output overlaid on the CT axial slices in Fig.~\ref{fig:covid-lesion}. Compared to the ground truth, the 3D lesion segmentation network tends to miss small GGO components as well as the pleura consolidations. Observing from the middle column of Fig.~\ref{fig:covid-lesion}, the synthesis augmented network (3D Network + Synthetic data) has higher sensitivity for such challenging regions without producing extra false positives.

We measured the severity of COVID-19 in each subject from predicted segmentation mask by DICE Similarity Coefficient (DSC), Percentage of Opacity (PO) and Percentage of High Opacity (PHO). The Percentage of Opacity is calculated as the total percent volume of the lung parenchyma that is affected by disease:
\begin{equation}
    PO = 100 \times \frac{\textrm{volume of predicted abnormalities}}{\textrm{volume of lung mask}}
\end{equation}
The Percentage of High Opacity is calculated as the total percentage volume of the lung parenchyma that is affected by severe disease i.e., high opacity regions including consolidation:
\begin{equation}
    PHO = 100 \times \frac{\textrm{volume of high opacity region}}{\textrm{volume of lung mask}}
\end{equation}
We measured these metrics on 100 COVID-19 positive and 100 control testing subjects to evaluate the ability of segmentation networks for predicting disease severity, which is summarized in Table~\ref{tbl:segmentation_result}. We evaluated 8 different abnormality segmentation strategies by comparing the following methods: the lung segmentation network finetuned on pneumonia images (Pneumonia Finetuned) vs the lung segmentation network trained on pneumonia images and synthetic images (Pneumonia Finetuned + Syn), the 2D segmentation network (2D) vs the 3D segmentation network (3D), and the segmentation network trained without synthetic images and the segmentation network trained with synthetic images (20\% of total training images). Using the three metrics, the user variability between different raters was estimated with 13 COVID-19 positive cases, which is shown in Table~\ref{tbl:user_variability}. 

\begin{table*}[ht]

\caption{DICE similarity coefficient and Pearson's correlation coefficient between predicted disease severity and measures derived from ground-truth for 100 COVID-19 positive and 100 control test cases.}
\centering
\resizebox{1\linewidth}{!}{
\begin{tabular}{l l l l l l}
\toprule
\multicolumn{3}{c}{Method} & DSC & Pearson's correlation on PO & Pearson's correlation on PHO \\
LungSeg & LesionSeg & 20\% Syn & & \\ 
\midrule
Pneumonia Finetuned & 2D & \xmark & $0.6232 \pm 0.1640$ & 0.9078 (p-value = \num{1.2e-76}) & 0.9063 (p-value = \num{5.4e-76}) \\ \hline
Pneumonia Finetuned & 2D & \cmark & $0.6347 \pm 0.1553$ & 0.9334 (p-value = \num{4.2e-90}) & 0.9235 (p-value = \num{2.5e-84}) \\ \hline
Pneumonia Finetuned &  3D & \xmark & $0.6565 \pm 0.1727$ & 0.9331 (p-value = \num{6.9e-60}) & 0.9099 (p-value = \num{1.3e-77}) \\ \hline
Pneumonia Finetuned &  3D & \cmark & $0.6915 \pm 0.1719$ & 0.9591 (p-value = \num{1.6e-110}) & 0.9218 (p-value = \num{1.9e-83}) \\ \hline
Pneumonia Finetuned+Syn &  2D & \xmark & $0.6348 \pm 0.1588$ & 0.9187 (p-value = \num{8.5e-82}) & 0.9120 (p-value = \num{1.4e-78}) \\ \hline
Pneumonia Finetuned+Syn &  2D & \cmark & $0.6450 \pm 0.1511$ & 0.9393 (p-value = \num{6.6e-94}) & 0.9270 (p-value = \num{2.7e-86}) \\ \hline
Pneumonia Finetuned+Syn &  3D & \xmark & $0.6786 \pm 0.1646$ & 0.9307 (p-value = \num{2.1e-88}) & 0.9317 (p-value = \num{4.9e-89}) \\ \hline
Pneumonia Finetuned+Syn &  3D & \cmark & $\boldsymbol{0.7064 \pm 0.1586}$ & $\boldsymbol{0.9613}$ (p-value = \num{8.7e-113}) & $\boldsymbol{0.9387}$ (p-value = \num{1.7e-93}) \\ \hline
\bottomrule
\end{tabular}}
\label{tbl:segmentation_result}
\end{table*}

\begin{table}[ht]
\caption{User variability between different readers. DICE similarity coefficient and Pearson's correlation coefficient (PCC) between two sets of annotated disease severity measures were used to estimate the user variability for 13 COVID-19 test cases.}
\centering
\begin{tabular}{c c c}
\toprule
DSC & PCC on PO & PCC on PHO  \\ \midrule
\multirow{2}{*}{$0.7132\pm0.1831$} & 0.9547 & 0.9702 \\ 
 & (p-value = \num{4.0e-7}) & (p-value = \num{4.0e-8}) \\ \hline
\bottomrule
\end{tabular}
\label{tbl:user_variability}
\end{table}

\section{Discussions \& Conclusions}
Rapid development of an AI system to evaluate CT imaging for COVID-19 is a challenging task for several reasons. The amount of standardized and curated CT image data from multiple centers is limited or restricted, the clinical protocols and indications for chest imaging in the disease are evolving as we speak and finally, there is an urgent need for a technological solution to ease the burden on healthcare systems in the middle of a pandemic. In this paper, we propose a solution to these challenges by introducing a framework to synthesize COVID-19 related tomographic patterns in CT images using GAN-based image inpainting and a 3D shape generation algorithm specific to COVID-19 patterns. In our work, we use a single abnormality label to annotate all COVID-19 patterns to train the synthesizer and the abnormality detection networks. Our solution therefore, not only augments data for training of algorithms, it also implicitly learns pneumonia patterns most commonly found in COVID-19, such as GGO, consolidation, and crazy paving pattern. 
\par The synthesizer is trained with 227 COVID-19 positive cases. We evaluate the impact of adding synthetic data to the lung and abnormality segmentation networks on a benchmark dataset of 100 COVID-19 positive patients and 100 control subjects. We evaluated the improvement in lung segmentation by a new metric called lesion inclusion rate or $LIR$, which measures the coverage of abnormalities in the lung mask, especially in locations that are common in COVID-19 such as the periphery. We find that addition of synthetic data improved the PIR by 6.02\%. Next, we evaluated the improvement of abnormality segmentation with the addition of synthetic data. We see that the DSC of the 2D network improved from 0.623 to 0.645 and the DSC of the 3D network improved from 0.657 to 0.706, which is comparable to the inter-user variability DSC ($0.7132\pm0.1831$). Finally, we also report that the Pearson's correlation coefficient between the ground truth and predicted metrics improved with networks using synthetic data. The PCC for the PO improved from 0.908 to 0.939 for the 2D network and 0.933 to 0.961 for the 3D network, which is comparable to the inter-user variability range ($PCC=0.957$). Likewise, the PCC for the PHO improved from 0.906 to 0.927 for the 2D network and 0.9099 to 0.9387 for the 3D network. 
We demonstrate in this paper that the addition of synthetic data improves the quality of lung segmentation by including the regions of high abnormality, which also translates to an improvement in abnormality segmentation. 

\textbf{Limitations}: At the moment we focused on the most prevalent patterns seen on chest CT images from COVID-19 confirmed patients, namely ground-glass opacities, consolidations and crazy paving patterns. Other abnormalities associated less frequently with chest CT images from COVID-19 patients such as atelectasis, interlobular septal thickening, pleural effusions and bronchiectasis are also present in our data set. However, we did not focus on synthesizing or validating our existing system on these specific findings.

\noindent\textbf{Disclaimer}: The concepts and information presented in this paper are based on research results that are not commercially available

\bibliographystyle{IEEEtran}
\bibliography{refs.bib}{}

\end{document}